\documentclass[a4paper,10pt]{article} 
\usepackage{amsmath}
\usepackage{graphicx}
\usepackage{subfigure}

\begin{document}



\title{Non-ideal memristors for a non-ideal world}

\author{Ella Gale$^{1,2}$\footnote{Current address: School of Experimental Psychology, University of Bristol, 12a Priory Road, Bristol, BS8 1TU. Email: ella.gale@bristol.ac.uk}\\
  1. Bristol Robotics Laboratory, Bristol, UK, BS16 1QY\\
  2. Department of Computer Science and Creative Technology,\\ University of the West of England, Bristol, UK, BS16 1QY}
  
\date{November 2014}

\maketitle   

\begin{abstract}
Memristors have pinched hysteresis loops in the $V-I$ plane. Ideal memristors are everywhere non-linear, cross at zero and are rotationally symmetric. In this paper we extend memristor theory to produce different types of non-ideality and find that: including a background current (such as an ionic current) moves the crossing point away from zero; including a degradation resistance (that increases with experimental time) leads to an asymmetry; modelling a low resistance filament in parallel describes triangular $V-I$ curves with a straight-line low resistance state. A novel measurement of hysteresis asymmetry was introduced based on hysteresis and it was found that which lobe was bigger depended on the size of the breaking current relative to the memristance. The hysteresis varied differently with each type of non-ideality, suggesting that measurements of several device I-V curves and calculation of these parameters could give an indication of the underlying mechanism.
\end{abstract}

\section{Introduction}

The memristor is a novel circuit element first proposed in 1971~\cite{14} and thought to be the missing fundamental circuit element that would relate charge, $q$, to magnetic flux, $\varphi$. The memristor could offer intriguing solutions to various technological problems such as low power computing, resilient electronics, neuromorphic computing due to its ability to keep a state without external power, possible resilience of that state to electromagnetic perturbation and brain-like combination of memory with processing in the same device and spiking properties. 

The original paper~\cite{14} defined a memristor device that is now-accepted as the ideal case. There are six properties that can be derived from this definition, and thus the ideal memristor: 
\begin{enumerate}
\item is two-terminal;
\item is a function of a single state variable (usually $w$ the boundary between TiO$_2$ and TiO$_{2-x}$);
\item relates $d\varphi = M dq$, with $M$ being the memristance, it is function between $V$ and $I$ that is non-linear everywhere (because $V$ and $I$ are the time-differentials of $\varphi$ and $q$ respectively);
\item has a rotationally symmetric pinched $V-I$ curve that resembles an infinity symbol rotated by 45 degrees; 
\item is passive (does not store or provide energy)
\item appears to cross at 0 in the $V-I$ plane (as a truly passive device cannot store energy so should have zero current at at zero voltage). 
\end{enumerate}


\begin{figure*}[htb]%
\subfigure[Curved Memristor]{%
\includegraphics*[width=.45\textwidth]{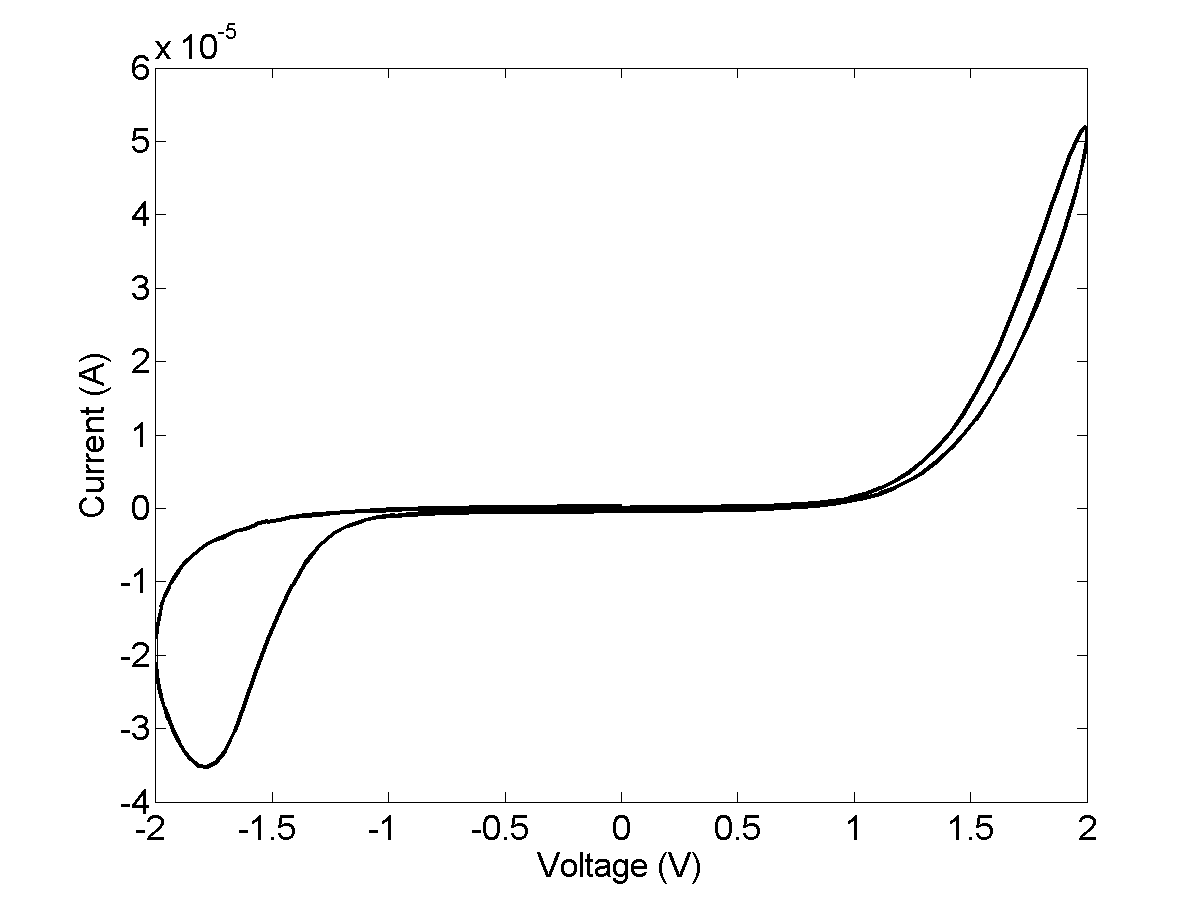}}\hfill
\subfigure[Filamentary Memristor]{%
\includegraphics*[width=.5\textwidth]{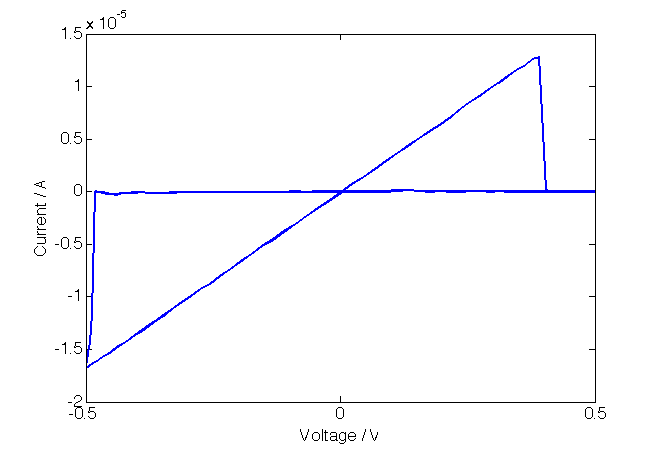}}\hfill%
\caption{Examples of experimentally-measured memristor curves for titanium dioxide sol-gel memristors}
\label{fig:Experimental}
\end{figure*}

Real-world memristors, such as those in figure~\ref{fig:Experimental} do not exactly resemble the original memristor definition (see the recent review~\cite{RevMemReRAM} for a discussion on the history of these definitions). As a result, several theoreticians have extended the concept of the memristor. Property 1 (two-terminal) was extended by the discovery of a three-terminal memristor~\cite{5,12}. Property 2 was expanded in 1976 with the introduction of the concept of a memristive system (or extended memristor~\cite{Chua8}), which could have more than one state variable~\cite{84}. Property three, memristors that had entirely non-linear resistance states, was contravened by many experimental observations (including~\cite{15,28} and the device in figure~\ref{fig:Experimental}b). Memristive systems allow the description of filamentary memristors, where a high resistance state that is linear in $V-I$ is allowed by associating a second state variable with the connection state of a filament~\cite{255}. The concept of an active memristor has been introduced~\cite{251} (a memristor that can store and/or output energy), which has proved useful in modelling living memristors~\cite{277} and which expands the description of memristors away from passivity as in property 5. Real memristor devices possess properties such as: non-zero crossing pinched hysteresis loops, or open curves~\cite{M1} (starting from~\cite{84}, these have been theoretically described from a physics point of view in~\cite{93}); off-set crossing pinched hysteresis loops (an example is given in~\cite{hystc}); and non-rotationally symmetric curves (seen in many experimental memristor system, see~\cite{RevMemReRAM} for a review). A forthcoming paper~\cite{Sah8} covers some of these forms of non-ideality from a circuit theoretic perspective, as applied to models of thermistors and neuronal ion channel memristors. In this paper we will deal with these three types of non-linearity in a theoretical manner using extensions to the memory conservation theory of memristance~\cite{F0c}. 

The memory-conservation theory of memristance~\cite{F0c} is the badly-named theory that describes memristors as a two-level system. The base level relates the magnetic flux, $\varphi_v$, associated with the drift of oxygen vacancies in a uniform magnetic field with the charge associated with those vacancies $q_v$. This is done by calculating the magnetic field associated with the vacancies, which gives an equation relating charge to flux, which is called the Chua memristance because it satisfies Chua's constitutive definition for the memristor (as given in~\cite{14}). This is a memristance as experienced by the oxygen vacancies, so we convert from vacancy-experienced resistance to electron-experienced resistance (called the memory function) using a single fitting parameter, which has been shown to fit experimental data~\cite{255}. This approach only models the doped (on) part of the device, so we use the definition of resistivity to describe the undoped (off) part of the device, the conservation function (the name comes from the requirement to conserve matter in the theory that this function satisfies) has a single fitting parameter, which has been fitted to experiment and shown to model the system well~\cite{255}. The memory and conservation functions are both memristive in that they will give memristor behaviour. Note that in the vanilla memory-conservation theory, we calculate only the electronic current and neglect the vacancy current itself (although vacancy resistance changes are included) as it was assumed to be small. 

In this paper we will present new extensions relating to two of the missing types of non-linearity: A. Non-zero crossing via off-set; B. rotational asymmetry. These results will be covered via extensions to the vanilla memory-conservation theory. These will be compared with a (simplified) filamentary memristor model. The effect of different parameters on the hysteresis contained in the pinched hysteresis curves will calculated. Our results are presented normalised to the maximum current so we can better compare the shapes of these curves which may exhibit variance across several orders of magnitude.

\section{Theoretical Methodology}

\subsection{Adding a second current}

The ion mobility of the vacancies, $\mu_v$, is slower than the ion mobility of the electrons, $\mu_e$, so drift velocity of the vacancies, $\vec{s}_v$ is slower than that of the electrons, $\vec{s}_e$. The time taken for the memristor short-term memory to dissipate is $\tau_{\infty}$, which is around 4s in our devices. The minimum time we can measure is $\tau_{0}$, which is limited by the maximum rate of measurement with the Keithley. 

The memory conservation theory does not include the ionic current as the assumption is made that the ionic current is negligible. There is evidence to suggest that this is not the case, including data from the plastic memristor~\cite{87} where the ionic current is often reported on the same or similar order as the electronic current, and the active memristor model of the slime mould which showed that the internal battery current was associated with a resistance at around 90\% of the starting measurement resistance~\cite{277}. Thus, we shall discuss an extension to the memory-conservation theory and demonstrate that a second current associated with the ionic charge can explain the non-zero crossing of the $I-V$ loops. 

The total current, $I$, is a sum of the electronic current, $i_e$ and ionic current, $i_v$:

\begin{equation}
I = i_e + i_v \, .
\end{equation}

We shall assume that the measurement rate is slow enough that all electronic lag due to the `inertia' of the electrons is dissipated -- not unreasonable given that our measurement frequency is around 1Hz, the size of the step-size that corresponds to this is $t_0$. As in~\cite{SpcJ}, the minimum current in $I(\tau_{\infty})$, the maximum measured is $I(t_0)$ (the first measurement made after the voltage is changed, in a perfect experimental set-up this would be $t_0=\delta$). The difference between these two values is $\Delta I$, given by $\Delta I = I(\tau_0)-I(\tau_{\infty})$. 

Thus, we can represent the total current as:

\begin{equation}
I = x_v \Delta I + I (\tau_{\infty}) \; . 
\end{equation} 

If $x_v>1$, the short term memory is exhausted. This is the part of a memristor's dynamics when the frequency is so slow that the device is locally acting like a resistor (in that locally between nearby voltages there is no real change in device resistance, the device may still be a memristor as it is possible for two branches of the memristor to act like resistors of different values with a discontinuous switch between them -- a device like this would still have a hysteresis, but it would have a non-linear relation between I and V in parts of the I-V loop. This case would be an ohmic resistive switch). 

If $x_v<1$, the short-term memory has an effect. 

We make the following assumption: ionic current after equilibration is negligible: $i_v (\tau_{\infty}) =0$. This implies that the current at this point is only due to the electrons: $I(\tau) = i_e$. 

We now artificially separate out the effect of the ionic-caused resistance change (as sampled by the electrons) and the ionic current. We define $r$ as the ratio of the total current at time $\tau_0$ due to the vacancy current, as $r = \frac{i_v(\tau_0)}{I(\tau_0)}$. And we simplify the situation by assuming that the difference between peak and equilibrated current does not change from step to step (not a bad assumption given the data in~\cite{SpcJ} if we keep to a small voltage range), i.e.: $\Delta I(t) \approx \Delta I(t-1)$. Note that we are using $\tau$ as the internal timescale for a current response of the system (which we expect to be related to $\vec{s}_{v}$) and $t$ is the running external time of the experiment, specifically the discretised measurement step.  

Thus, for step $t$, we can write
\begin{equation} 
I(t) = I(\tau_{\infty} |_t + x_V r \Delta I |_{t-1})
\end{equation}

where the vertical bar means the expression is evaluated at that external time, i.e. $I(\tau_{\infty} |_t)$ would be the current due only to the decay of the short-term memory as expected at time $t$ after stimulation. The simulation we do is:

\begin{equation}
I(t) = i_e(t) + x_v r i_e(t-1) \; ,
\end{equation}

where $i_e$ is calculated from the memory-conservation theory and we have taken the assumption that $I(t)$ is only due to $i_e$ (an assumption of `vanilla' mem-con theory). Essentially, the second term adds a second charge carrier which is lagging the electronic current more slowly.  

Analytically, this neatly explains where the current at $V=0$ comes from: 

\begin{equation}
I(V=0) = i_e(V=0) + x_v r i_e(\pm V_{stepsize}) \;,
\end{equation}

the second term is still feeling the effects of the previous non-zero voltage, which is tending to zero, but slow enough that current is still flowing one time-step later. In this equation $V_{step-size}$ is the size of voltage steps. This also explains why the current is negative at the end of the second quadrant, as the memristor goes from $V_{step-size} \rightarrow 0V$, the difference $\Delta V$ is negative and as in~\cite{UCNC} we know that a negative $\Delta V$ leads to a negative current impulse, $r i_e$, which we are sampling $t$ seconds later (which is accounted for by $x$). 

\subsection{Adding in a second increasing resistance}

The material in our devices is a thin-film semi-conductor, thus the field across it is huge. This leads to changes in the structure of the semiconductor material such as Joule heating, filamentary fusing and anti-fusing, phase changes and so on (see~\cite{RevMemReRAM} for a full list of possibilities and experimental evidence for these mechanisms). Here we will not model a specific mechanism but instead assume that there is a second resistance associated with a degradation of the device.

We shall define $R_{\mathrm{total}}$ as the total resistance of the system run with only the vanilla memory-conservation theory. We define $R_2$ as the resistance associated with the degradation of the device during testing. We want $R_2$ to reach its maximum value during the $V-I$ cycle so we define the update change in $R_2$ as $\Delta R_2$ given by:

\begin{equation}
\Delta R_2 = x_r \frac{1}{n} \mathrm{Max}[R_{\mathrm{total}}] \;
\end{equation}

where $n$ is the number of time-steps in a cycle, Max[$y$] is a function to take the maximum value of $x$, and $x_r$ is a multiplier which we can vary to investigate the effect of the size of $R_2$. 

The update code for the total resistance is then simply:
\begin{equation}
R_2(j) = R_2(j-1) + \Delta R_2
R_{\mathrm{total}}(j) = R_{\mathrm{total}}(j) + \Delta R_2
\end{equation}

\subsection{Adding in a Filament}

Devices which switch to an ohmic low resistance state, believed to be due to the presence of a filament of a either a higher-conducting semi-conductor phase or metallic phase which connects. These devices have been known for years in the field of ReRAM (see~\cite{155}) and are not ideal memristors, but do fit the definition of memristive systems. To extend the memory-conservation theory, a filament resistance, $R_\mathrm{fil}$, was added in parallel to the ideal memristor, along with a switch (theoretically represented by a Heaviside function) which closed and allowed current to flow through the $R_{\mathrm{Fil}}$ when the filament reached the end of the device. An equivalent circuit is shown in~\ref{fig:circuit}. 

Thus, the extended system resistance, $R$, is given by
\begin{equation}
R = \frac{1}
{
	\frac{1}
	{
		R_{\mathrm{total}}
	} 
	+ 2 H \left( w - D \right) \frac{1}
	{
		R_{\mathrm{Fil}}
	}
	},
\label{eq:Switch}
\end{equation}
where $H$ is the Heaviside function as implemented in MatLab which gives the following values (for $x$ where $x$ is the element of the positive reals): 
\begin{itemize}
 \item $\: H(-x) = 0$
 \item $\: H(+x) = 1$
 \item $\: H(0) = \frac{1}{2} \; .$ 
\end{itemize}

As the resistance of this filament is much lower than that of rest of the device most the current goes through it, leading to a low resistance state one or more orders of magnitude higher than the high resistance state. Previous work~\cite{255} took into account the fractal nature of the filament, here to add comparison with the other data, $R_{\mathrm{Fil}}$ is presented in units of of Min[$R_{\mathrm{total}}$], which is the minimum resistance of an equivalent ideal memristor run under the same conditions.

\subsection{Calculation of the Asymmetry Metric}

The memristor plot is split into 4 branches: 1: $0 < V < +V_{\mathrm{max}}$; 2: $+V_{\mathrm{max}} < V < 0$; 3: $0 < V < -V_{\mathrm{max}}$; 4: $-V_{\mathrm{max}} < v < 0$. The hysteresis, $H$, scaled hysteresis, $\bar{H}$ (scaled relative to the resistance of a resistor of device starting resistance, $R_0$) was calculated from the work done by the device over each branch of the plot as in~\cite{G1}. The asymmetry metric, $A$, was calculated from the difference of the hysteresis of the positive lobe and the negative lobe as:

\begin{equation}
A = (W_2-W_1)-(W_3-W_4)
\end{equation}
where $W_x$ is the work associated with branch $x$.

\subsection{Simulation details}

All simulations were done in MatLab using reduced units as in the simulation in~\cite{15,M0}. All $V-I$ plots were scaled so the maximum current was 1. Except where stated otherwise, we used 160 timesteps and a frequency of 0.4$\omega_0$, this ensures that the $w$ can not move out of its allowed range ($0<w<D$) and means that we do not require memory functions. 

\section{Results}

\subsection{Non-zero Crossing}

\begin{figure*}[htb]%
\subfigure[Zero-crossing V-I Curve]{%
\includegraphics*[width=.5\textwidth]{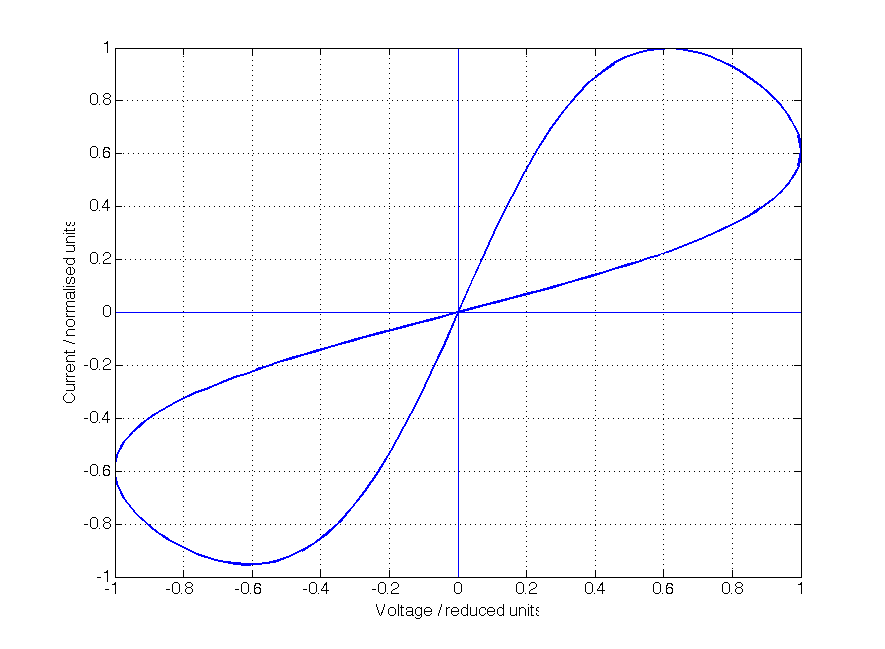}}\hfill
\subfigure[Off-set crossing point]{%
\includegraphics*[width=.5\textwidth]{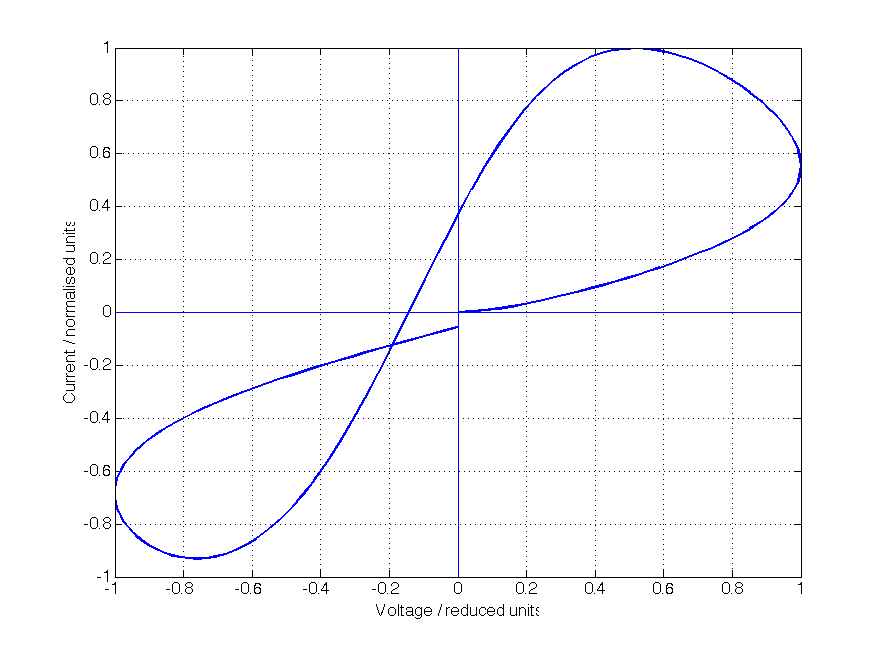}}\hfill%
\caption{The effect of changing the product $x_v r$. In a. $x_v r = 0$ which is equivalent to vanilla theory, in b. $x_v r = 0.8$ gives an offset of -0.2 V}
\label{fig:asym}
\end{figure*}

\begin{figure*}[htb]%
\subfigure[$x_v r$ from 0 to 1 in jumps of 0.1]{%
\includegraphics*[width=.5\textwidth]{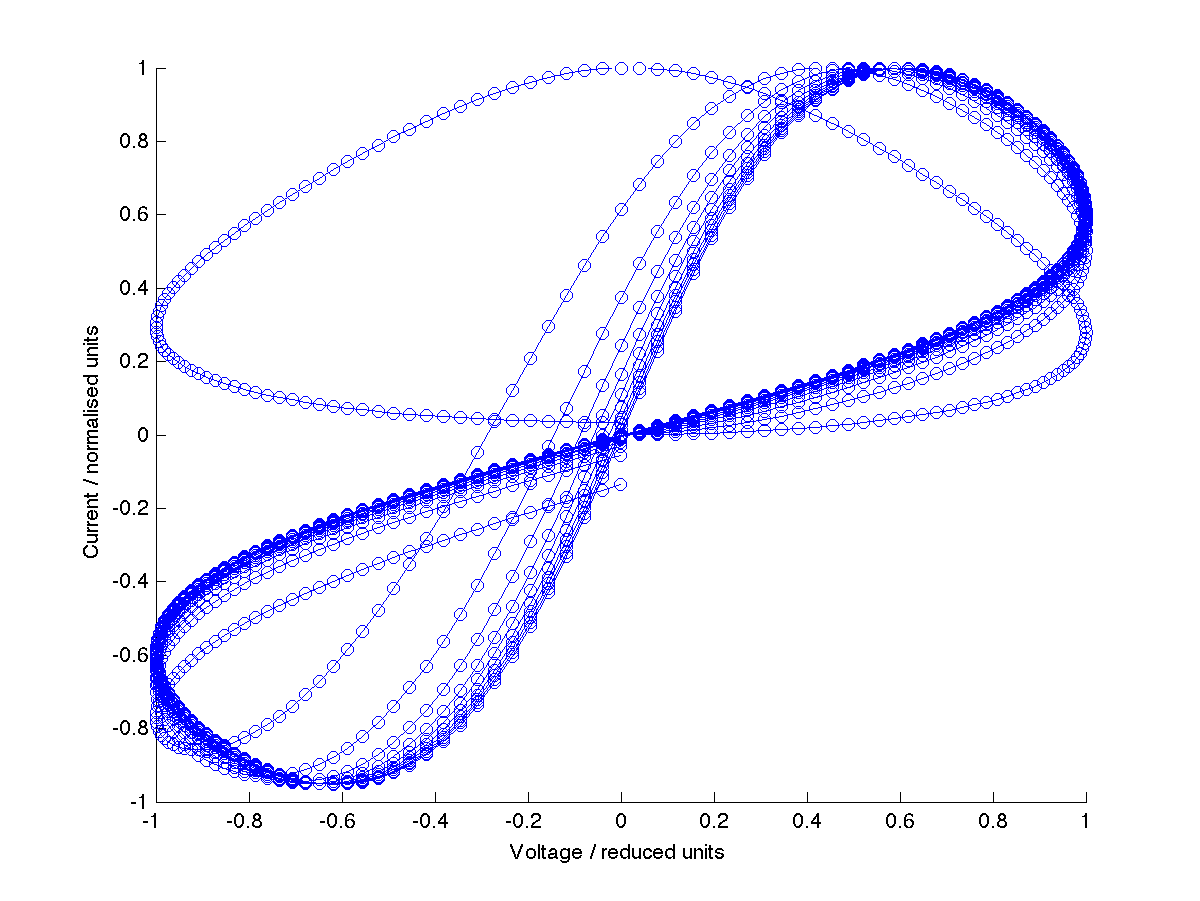}}\hfill
\subfigure[$x_v r$ from 0.9 to 1 in jumps of 0.01]{%
\includegraphics*[width=.5\textwidth]{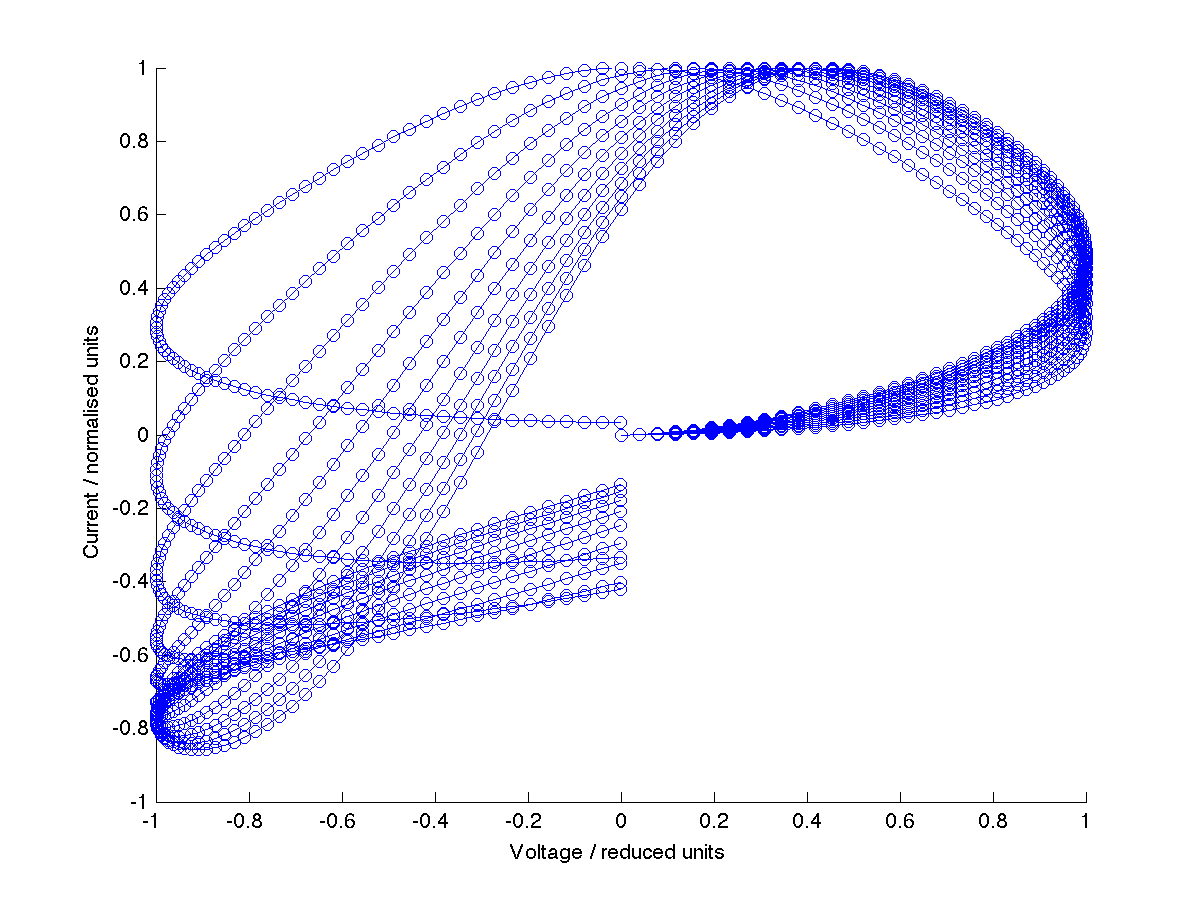}}\hfill%
\caption{The effect of changing the value of $x_v r$ from 0 to 1 in jumps of 0.1 (this is for one loop, with repeated loops the curve would join up at zero V and give a trace similar to that observed in experiments).}
\label{fig:asym}
\end{figure*}

Results are shown in figure~\ref{fig:asym}, we can see that having a product of $x_v r =0.8$ moves the crossing point to -0.2V, this is because the ionic current lags to electronic current so goes to zero after device has gone through 0 applied voltage. Also, the asymmetry of the right-hand figure is 11 times larger than the left hand one (the asymmetry is not zero for the ideal curve due to the small number of time-steps used for these simulations, although it should be analytically zero for a smooth curve and tends to zero as the number of simulation timesteps is increased). 

As we can see in figure~\ref{fig:asym}, by changing the value of the product $x_v r$ the crossing point moves further to the negative due to the lag of the vacancy current. For the special case of $x_v r = 1$, which is the condition where the vacancy current is equal to the electronic current the previous step, we get a strange open curve that is nowhere negative and resembles a mushroom. This is not necessarily a physical system for a memristor, in fact, using this update rule on an ohmic resistor gives a perfect circle in $V-I$ space which we associate with a capacitor. 

\begin{figure}[htbp]%
\includegraphics*[width=\linewidth,height=\linewidth]{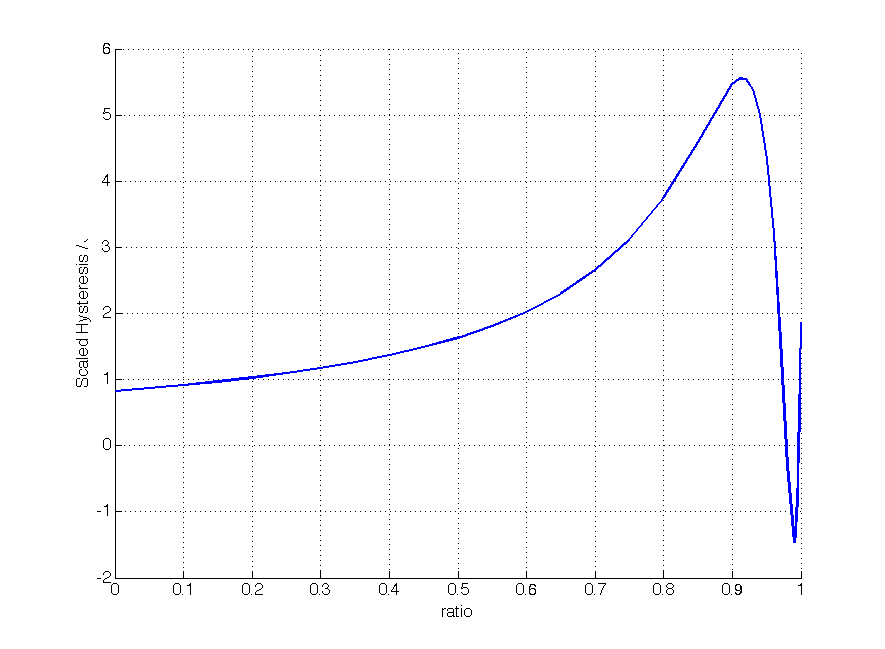}
\caption{%
  How the scaled hysteresis varies with ratio $r$ ($x_v$ was set equal to 1).}
\label{fig:2Hbar}
\end{figure}

Figure~\ref{fig:2Hbar} shows how the scaled hysteresis changes with $x_v r$. This graph was evaluated at a different rate over the range of $r$ with simulations done at every 0.01 increment between 0.9 and 1 to capture the full range of behaviour. We see that the maximum hysteresis is seen at around $x_v r =0.9$.

\subsection{Non-Rotationally Symmetric}

\begin{figure*}[htb]%
\subfigure[Experiment A]{%
\includegraphics*[width=.5\textwidth]{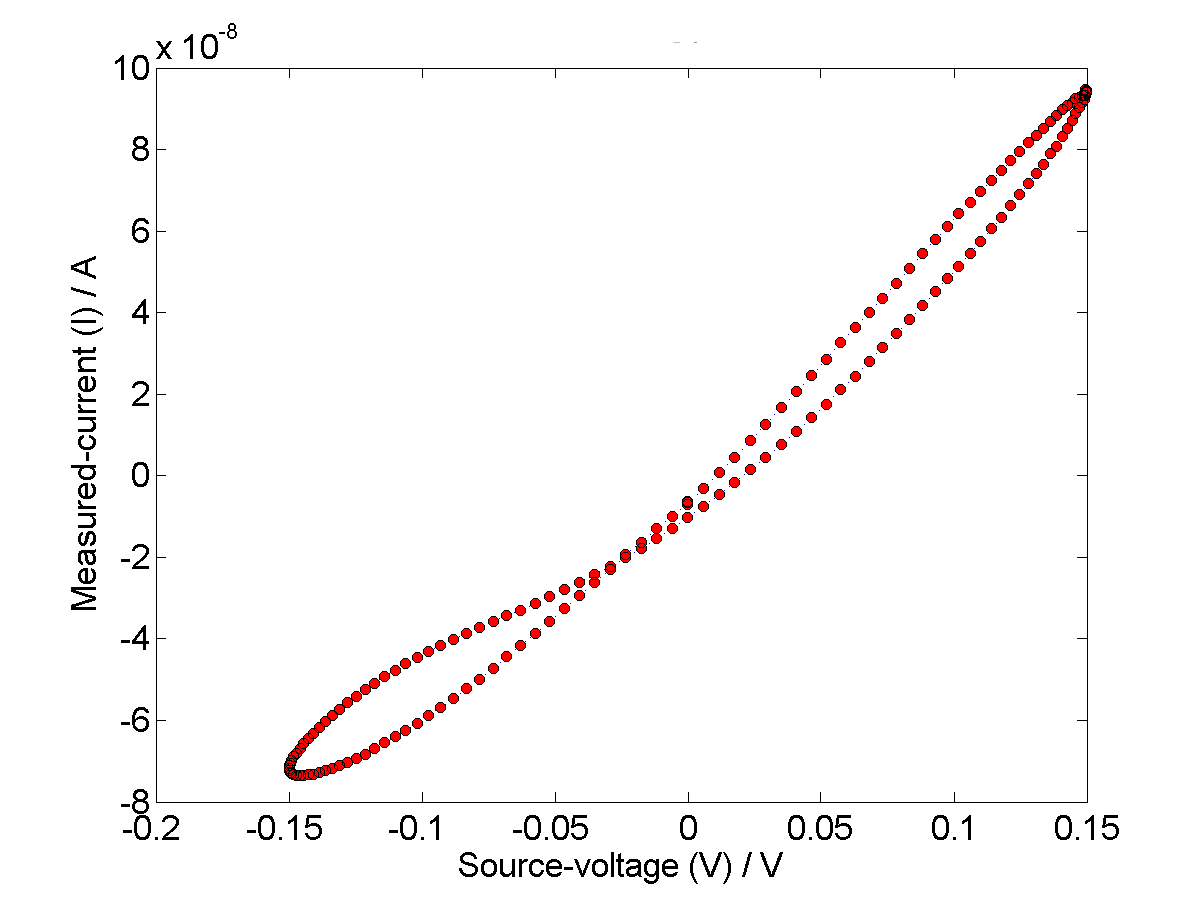}}\hfill
\subfigure[Experiment B]{%
\includegraphics*[width=.5\textwidth]{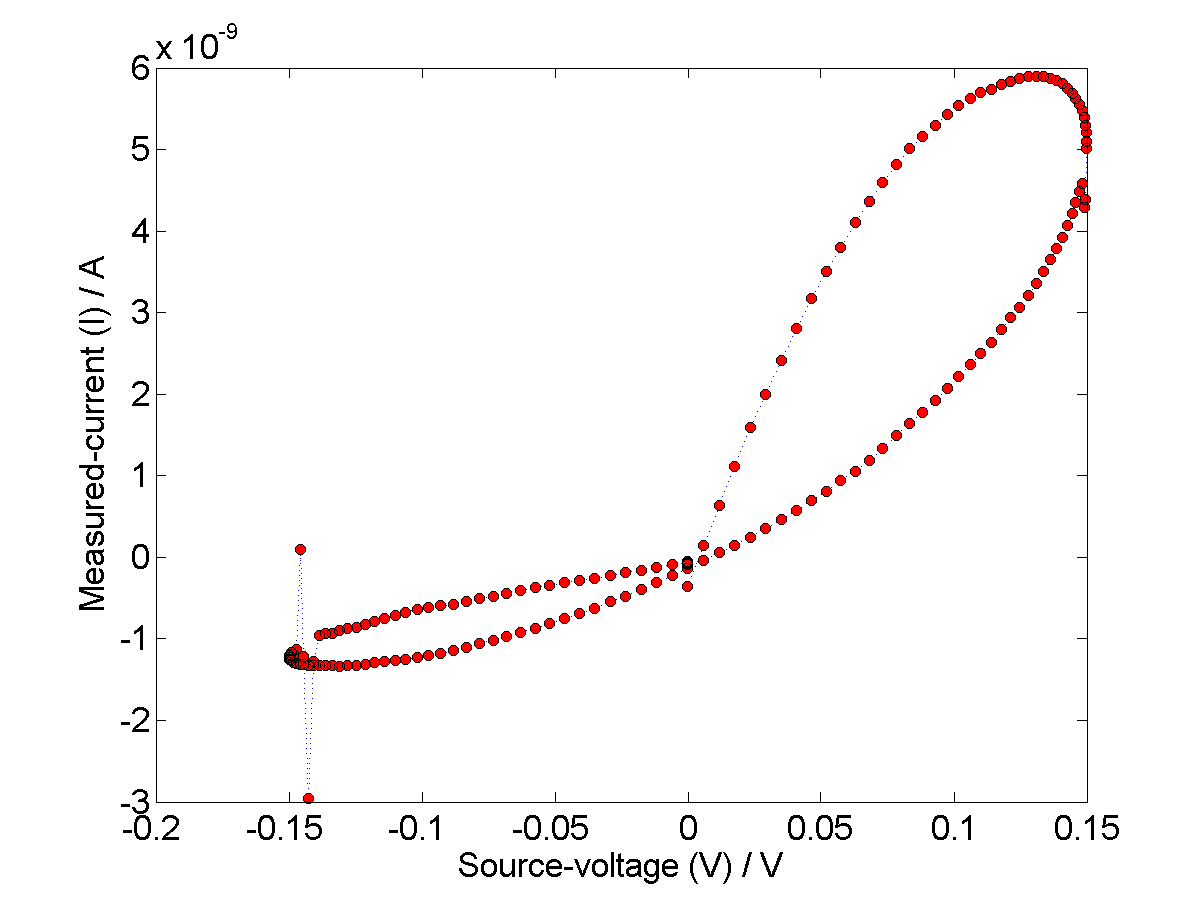}}\hfill%
\caption{Examples of (living) memristors that lack rotational symmetry, taken from~\cite{277}.}
\label{fig:Slime}
\end{figure*}

\begin{figure*}[htb]%
\subfigure[$R_2 < R_{\mathrm{total}}$]{%
\includegraphics*[width=.5\textwidth]{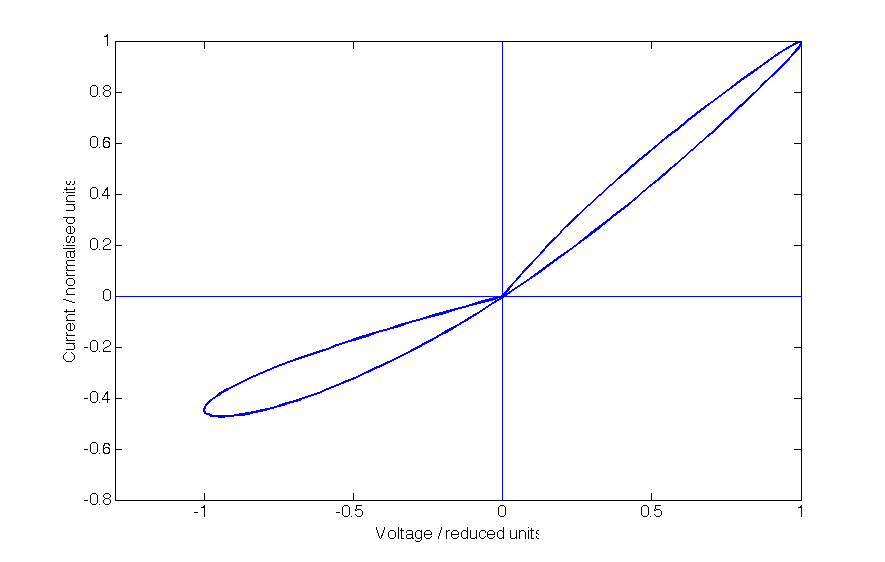}}\hfill
\subfigure[$R_2 > R_{\mathrm{total}}$]{%
\includegraphics*[width=.5\textwidth]{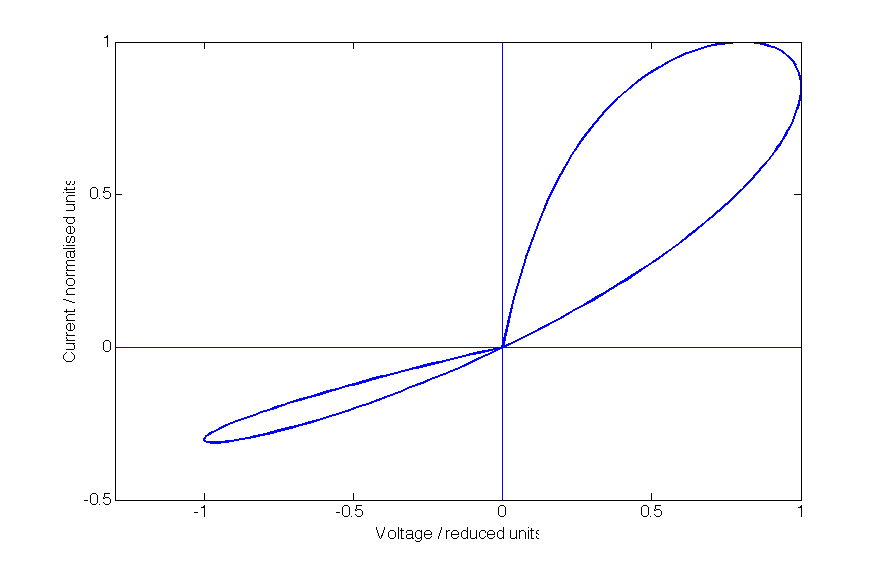}}\hfill%
\caption{Examples of behaviour observed with different values of device degradation resistance (remember that the currents are normalised so a larger $R_2$ results in a more extreme change between the positive and negative lobes).}
\label{fig:asym2}
\end{figure*}

Figure~\ref{fig:Slime} shows a few example experiments of non-rotationally symmetrical memristors. These examples are taken from biological memristors (electrical measurements of \textit{Physarum Polycephalum} cytoplasm~\cite{277}). Figure~\ref{asym2} shows example of different memristor curves that closely resemble the experimental data in figure~\ref{fig:Slime}, figure~\ref{fig:asym2}a has an $R_2$ of $0.85\times R_{\mathrm{total}}$ and figure~\ref{fig:asym2}b has an $R_2$ of $4.45\times R_{\mathrm{total}}$. As the currents are normalised, the effect of a larger $R_2$ is a greater asymmetry in the curve; we would also get a smaller overall current. Interestingly, the relative lobe sizes changes dependent on whether the degradation resistance is larger than Max($R_{\mathrm{total}}$); positive lobe has a larger hysteresis, or smaller than Max($R_{\mathrm{total}}$) where the negative lobe has a larger resistance. 

The fact that this model produces qualitative behaviour of the correct form suggests that the biological memristors are being affected by testing, with an increase in resistance. This fits with observation made during the testing process that the amount of cytosol in the system decreased over the days of testing (the organism moved away to explore), leaving behind the gel outer-layer, which was observed to have a high resistance. Similar curves are also often seen in semi-conductor devices, and it is expected that these are also due to material changes, primarily due to internal nanoscale heating effects caused by testing.

\begin{figure}[htbp]%
\includegraphics*[width=\linewidth,height=\linewidth]{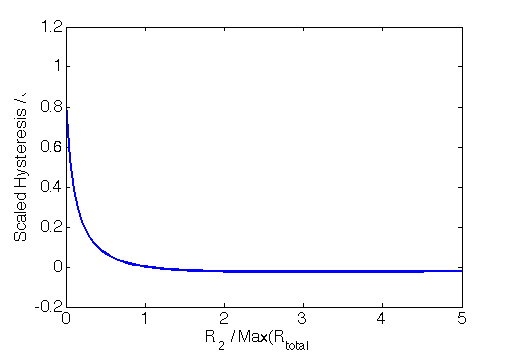}
\caption{%
  How the scaled hysteresis varies with the addition of a degradation resistance, $R_2$.}
\label{fig:2cHbar}
\end{figure}

The effect of the degradation resistance on the hysteresis is shown in figure~\ref{fig:2cHbar}. The hysteresis decreases with increasing $R_2$ as we have added an extra resistance to the system. The change from positive to negative hysteresis values is due to the change in which lobe is larger. The asymmetry metric is shown in figure~\ref{fig:2cHbar}. As the degradation resistance is relative to the maximum total resistance for a ideal device and changes linearly with time (whereas the current changes non-linearly with voltage, which changes sinusoidally with time), the asymmetry behaviour is not simple. The system changes from having a positive lobe larger than a negative at around $R_2 = 0.3\times \mathrm{Max}(R_{\mathrm{total}})$ and reaches a maximum asymmetry at around 1.2. 

\begin{figure}[t]%
\includegraphics*[width=\linewidth,height=\linewidth]{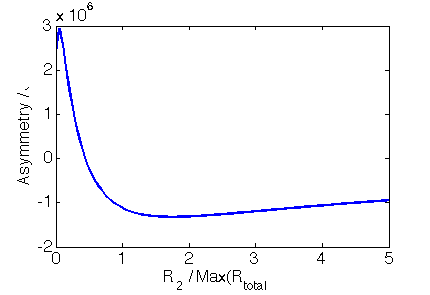}
\caption{%
  The change in asymmetry with increasing degradation resistance.
    }
\label{fig:2cAsym}
\end{figure}


\section{The effect of switching }

The filamentary model has two parameters that we can change, the switching point, $w$, and the resistance of the filament, $R_{\mathrm{Fil}}$. For the ideal memristor examples $w$ varies between 0.1 and 0.9.

\begin{figure}[htbp]%
\includegraphics*[width=\linewidth,height=\linewidth]{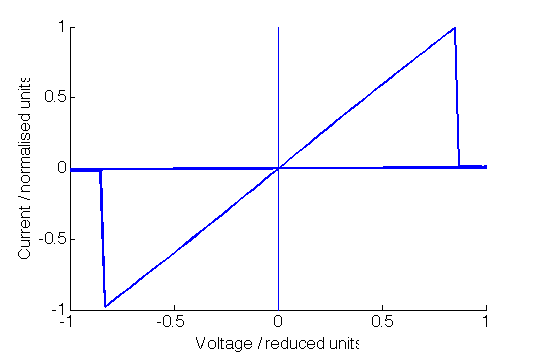}
\caption{%
  Example which switches when w=0.7 and the filament divisor of 1 }
\label{fig:FilBoth}
\end{figure}

\begin{figure}[htbp]%
\includegraphics*[width=\linewidth,height=\linewidth]{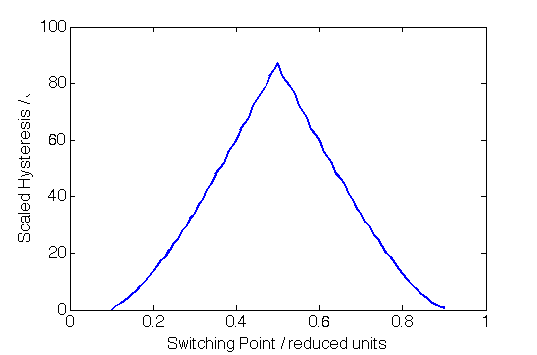}
\caption{%
  How the scaled hysteresis varies with switch point. }
\label{fig:SwitchHBar}
\end{figure}

This data came from simulations with twice the number of steps as using 160 steps gave results with the same hysteresis value, increasing time granularity smoothed that out and we expect that increasing step number will smooth the curve further. An example of the device that switches when $w=0.7$ with a $R_{\mathrm{fil}}=10$, (i.e. the filament resistance is an order of magnitude above the minimum resistance of the corresponding vanilla memristor) is given in figure~\ref{fig:FilBoth}; the dynamics are similar to those seen in~\cite{15}. The hysteresis change is shown in figure~\ref{fig:SwitchHBar}. This curve was ran at twice the step size of the others to smooth out the function. The hysteresis scaled increased with crossing point up to a maximum of 0.5 which corresponds to the entirety of the lower magnitude arm of each lobe being in the higher resistance state and the higher magnitude arm of each lobe being in the lower resistance state. The trend is not linear. 

\begin{figure}[htbp]%
\includegraphics*[width=\linewidth,height=\linewidth]{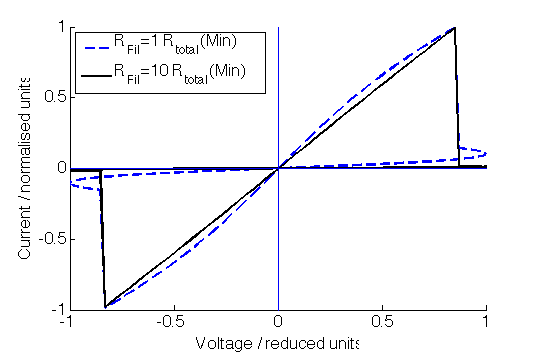}
\caption{%
  Two example values of $R_{\mathrm{Fil}}$}
\label{fig:EffectOfrFil}
\end{figure}

\begin{figure}[htbp]%
\includegraphics*[width=\linewidth,height=\linewidth]{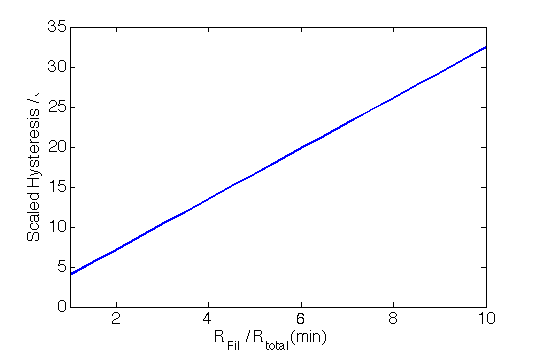}
\caption{%
  The hysteresis scales linearly with $R_{\mathrm{fil}}$}
\label{fig:HBarrFil}
\end{figure}

We can also change the filamentary resistance values as is shown in figure~\ref{fig:EffectOfrFil}; the blue line shows the special case when $R_{\mathrm{Fil}}$ and we can see that non-linearities from the background bulk vacancy movement are apparent in the $I-V$ curve. If we don't see that in experiments, it means that the filament is more than a order of magnitude above the bulk memristance. The hysteresis (not shown) and scaled hysteresis (as shown in figure~\ref{fig:HBarrFil} both vary linearly with increasing filamentary memristance.


\section{Conclusions}

In this paper, we have presented two novel extensions to the memory-conservation theory of memristance that take into account a vacancy/background current or a resistance that increases linearly with time (such as a resistance associated with device degradation), although the extensions are general and could be applied to any other memristor theory.

We found that a second vacancy current was sufficient to account for offsetting the pinched hysteresis loop crossing point from zero. This theoretical result complements the experimental findings in~\cite{292}, and is the extension to memristor theory that authors of that paper requested. 

We have demonstrated significant rotational asymmetry in pinched hysteresis loops can be introduced by adding in a degradation voltage and that options for which loop was larger are well explained by the value of the degradation resistance. We introduced a novel metric for measuring hysteresis loop asymmetry which works well as a data analysis approach. The results presented here suggest that changing the timescale of the measurement ought to effect the amount of degradation experienced and the resulting asymmetry in the $I-V$ curves. 

Finally, we undertook a similar analysis of the filamentary extension to the memory-conservation theory of memristance and found that hysteresis increased linearly with filament conduction. 

A comparison of the hysteresis graphs presented here suggests that it is possible to elucidate the experimental mechanism from I-V data. An I-V offset is related to a secondary current, either a vacancy current or a `nanobattery', finding the point at which the curve crosses zero gives information about the time-scale of this current. Asymmetric I-V curves are due to a degradation resistance. If the third quadrant hysteresis is larger than the first, the then degradation resistance is smaller than the resistance range of the memristive part of the device; whereas if the first quadrant hysteresis is larger than the third, the degradation resistance is larger than the memristive response. It was already known that an ohmic low resistance state is indicative of a filament. However, the dominant mechanism can be found by looking at how the values of hysteresis change with against $x_r$, $w$, a $R_2$ (all of which can be approximated based on V-I measurements) for several different devices or runs.

\section*{Acknowledgement}
E. Gale would like to thank Oliver Matthews, Ben de Lacy Costello and Andrew Adamatzky for support.

\clearpage

%
\bibliographystyle{pss}
\bibliography{UWELit}
%



\end{document}